\documentclass{vospwks2007}

\usepackage{graphicx}

\title{The Variable Star One-shot Project, and its little child: Wikimbad.}

\author[1,2]{Foellmi C\'edric}
\affil[1]{Laboratoire d'AstrOphysique de Grenoble, 414 Rue de la Piscine, 38400 Saint-Martin d'H\`eres, France}
\affil[2]{European Southern Observatory, 3107 Alonso de Cordova, casilla 19, Santiago, Chile}
\author[3]{T. H. Dall}
\affil[3]{Gemini Observatory, 670 N. A'ohoku Pl., Hilo, HI 96720, U.S.A.}
\author[4]{J.~Pritchard}
\affil[4]{European Southern Observatory, Karl Schwarzschild-Str. 2, D-85748 Garching bei M\"unchen, Germany}
\author[5]{C.~Allende Prieto}
\affil[5]{McDonald Observatory and Department of Astronomy, The University of Texas, Austin, TX 78712-1083, U.S.A.}
\author[6]{H.~Bruntt}
\affil[6]{School of Physics A28, University of Sydney, 2006 NSW, Australia}
\author[7]{P.~J.~Amado}
\affil[7]{Universidad de Granada-IAA(CSIC), P.O. Box 3004, 18080 Granada, Spain}
\author[8]{T.~Arentoft}
\affil[8]{Department of Physics and Astronomy, University of Aarhus, 8000 Aarhus C, Denmark}
\author[9]{M.~Baes}
\affil[9]{Sterrenkundig Observatorium, Universiteit Gent, Krijgslaan 281 S9, 9000 Gent, Belgium}
\author[10]{E.~Depagne}
\affil[10]{Pontificia Universidad Catolica de Chile, Vicuna Mackenna 4860, Santiago de Chile, Chile}
\author[11]{M.~Fernandez}
\affil[11]{Instituto de Astrof\'{\i}sica de Andaluc\'{\i}a, Camino Bajo de Hu\'etor 50, 18008 Granada, Spain}
\author[2]{V.~D.~Ivanov}
\author[5]{L.~Koesterke}
\author[2]{L.~Monaco}
\author[2]{K.~O'Brien}
\author[12]{L.~M.~Sarro}
\affil[12]{Department of Artificial Intelligence, ETSI Inform\'atica, Juan del Rosal 16, 28040 Madrid, Spain}
\author[2]{I.~Saviane}
\author[2]{J.~Scharwaechter}
\author[2]{L.~Schmidtobreick}
\author[2]{O.~Sch\"utz}
\author[13]{A.~Seifahrt}
\affil[13]{Astrophysikalisches Institut und Universit\"ats-sternwarte Jena, Schillerg\"assen 2, 07745 Jena, Germany}
\author[2]{F.~Selman}
\author[2]{M.~Stefanon}
\author[2]{M.~Sterzik}

\begin{document}


\maketitle

\begin{abstract}
The Variable Star One-shot Project (VSOP) aimed at providing to the world-wide stellar community the necessary one-shot spectrum of unstudied variable stars, too often classified as such by an analysis of photometric data only. The VSOP has established an new kind of observational model, where all steps from observations to spectral analysis, are automatized (or are underway to be fully automatized). The project is centralized on a collaborative wiki website. The VSOP operational model is very successful, data is continously flowing and being analyszed, and VSOP is now a worldwide open collaboration of people with very different and complementary skills and expertise. The idea of a central wiki website has been extended by one of us to propose a new service to the whole astronomical community, called Wikimbad. Wikimbad is an open wiki website aimed at collecting, organizing and making publicly available all kind of reduced and published astronomical data. Its strengths and a comparison with the Virtual Observatory are discussed.
\end{abstract}

\section{The VSOP Project}

There are more than 38,000 known variable stars listed in the latest edition of the General Catalog of Variable Stars \citep[GCVS;][]{Kholopov-etal-1998}. Almost 4,000 of these have no spectral type assigned and nearly 2,000 are listed with an uncertain variability type, often because of lack of spectroscopic characterisation.  The rate of discovery of new variables is currently around 500 per year \citep{Kazarovets-etal-2003}. About half of these newly identified variables have unknown variability type and most of them have no published spectral type. Moreover, many variables have disagreeing designations between different authors and even between different catalogs. In addition, binarity is rarely detectable unquestionably by photometric data alone. Finally, many designations are taken at face value without questioning the reliability. This unreliability is a major obstruction to many individual studies, and would often require only one ``snapshot'' spectrum to achieve a major improvement, even if a single shot spectrum would not always be sufficient to reveal binarity or transient phenomena. 
                   
Motivated by the situation outlined above, we have initiated a new kind of program, called the {\it Variable Star One-shot Project} (VSOP) aimed at collecting these one-shot spectra over a large number of variable stars. The goals of VSOP are:
\begin{enumerate}
\item To obtain the first spectroscopy of all unstudied variable stars.
\item To revise spectral and variability types, and when possible, the (spectroscopic) binary status.
\item To provide high-level scientific output such as Cross-Correlation function and stellar parameters such as gravity and effective temperature.
\item To automatize as much as possible the observational process: observations, reduction, binarity resolution, stellar parameters determination and ultimately variability determination.
\item To make the data available as automatically as possible, facilitating additional science.
\item To generate serendipitous discoveries by the sheer amount of data that will fuel future research.
\end{enumerate}

\begin{figure*}
\centering
\includegraphics[width=0.4\linewidth]{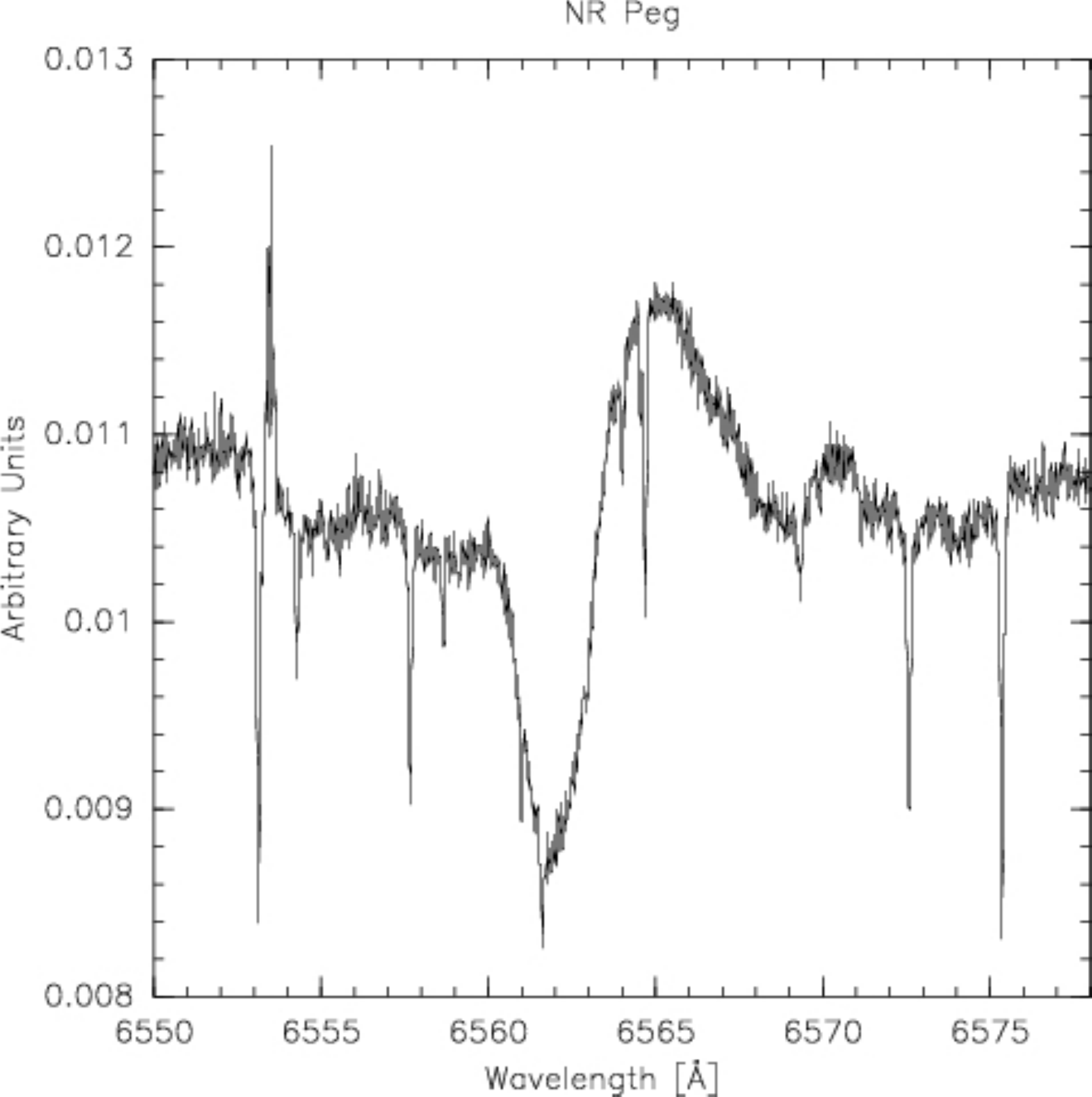}
\includegraphics[width=0.4\linewidth]{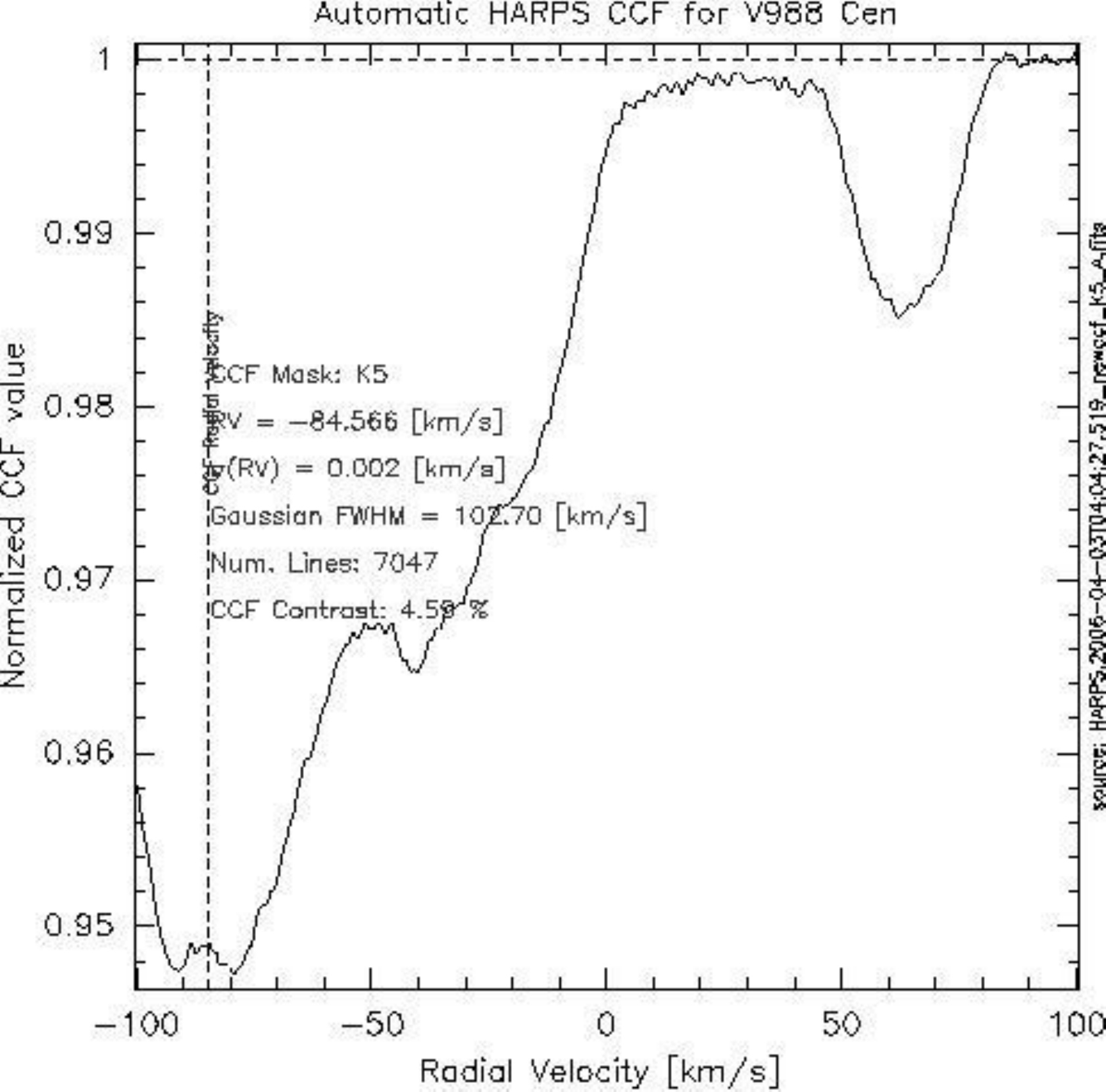}
\caption{Examples of VSOP products. Left: The HARPS spectrum of NR Peg around H$\alpha$, showing its P Cygni profile. Right: The Cross-Correlation function of V988 Cen revealing the complex binarity of this target.}
\label{ccfs}
\end{figure*}

Morover, a stellar spectrum is a rich source of information, and there are certainly many other scientifically interesting studies one could perform using our data. We choose, rather than to sit on the data until we may find time for additional studies, to make the data easily available to the general community, in the hope that somebody else will be able to do additional science with the data.  This way, the science output is maximised. This idea has been generalized with the Wikimbad project (see below).

Another aspect that contributes to the science efficiency, is serendipity. The VSOP observations
are targeting poorly studied variable stars, many of which are exhibiting poorly studied phenomena. We thus expect to obtain by chance data that either merit follow-up in-depth studies, or sheds light on some hitherto obscured phenomenon. Much of this work may naturally be done by groups not affiliated with VSOP.

Because many of today's VSOP astronomers are current or past support astronomers in the La Silla Observatory, VSOP was originally conceived as an ESO observatory project aimed at providing observations with loose weather and pointing constraints, with the aim of increasing observing efficiency during periods when other programmes with stronger constraints on airmass, seeing, and/or transparency cannot be carried out.  Given the all-sky coverage, the loose constraints, and the large scope of the project, VSOP is an excellent example of a perfect filler programme, which will be extended to other observatories in the near future \citep[see][]{Dall-etal-2007}.

\section{The VSOP Wiki Database}

All the information about the VSOP targets are stored in a wiki\footnote{Quickly-speaking, a wiki website is a website where pages can be directly edited from within the browser.} website located at  {\tt http://vsop.sc.eso.org}, from where the reduced data of this First Data Release \citep{Dall-etal-2007} can be freely accessed.

For the organization of information, we have chosen the MediaWiki software, developed for the open and free on-line encyclopedia Wikipedia\footnote{{\tt http://www.wikipedia.org}}. This ensure to have a software capable of supporting a reliable and extendable website where all VSOP members can contribute easily from their own daily workplace. This is of growing importance given the distribution of VSOP members around the world, as evidenced by the list of affiliations for this paper. 

The MediaWiki software is based on the article/discussion wiki philosophy, which means that to each  article page there is an associated discussion page. For VSOP we have extended the software to make the discussion pages restricted to VSOP members only, while the article pages are reserved for already published results, freely accessible to anybody. Thus, each star has a dedicated article page, where basic informations (coordinates, magnitude, link to SIMBAD, finding charts, old variability and spectral types -- when available) are provided. Also, the  observation details are described as well as the analysis, its results, a list of references, catalogues and download links to plots of the spectra as well as to all the reduced data products: Cross-Correlation Function (CCF) and wavelength calibrated one-dimensional spectrum). The use of MediaWiki is also motivated by the fact that it automatically keep tracks of the history of every pages, providing who and when made what modifications, and has "live" up-to-date page categories.

Finally, a major feature is that wiki websites using MediaWiki are easily scriptable. We have thus developed a VSOP-dedicated software module written in Python (another one in Perl is in developement) which makes the development of scripts dedicated to VSOP pages much easier. These robot scripts can then update a large amount of repetitive information, or collect the results of given subcategories of stars. 

\begin{figure*}
\centering
\includegraphics[width=0.6\linewidth]{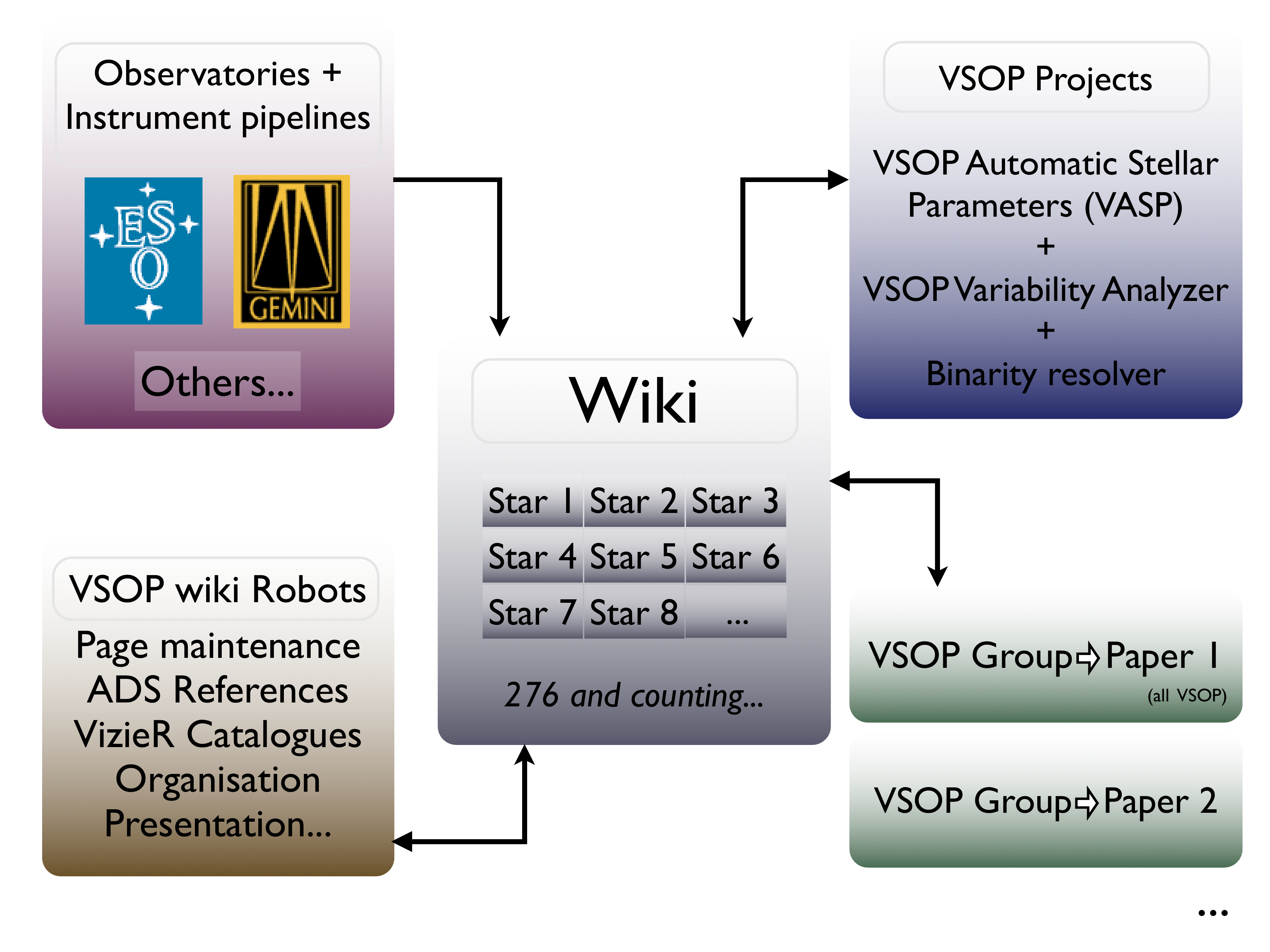}
\caption{The V.S.O.P. work model. Data is coming from the observatories, where dedicated pipelines take care of reducing the data, and automaticaly transferred to the VSOP machine at ESO Vitacura. VSOP "robots" (i.e. scripts) take care of the maintaining the wiki website up-to-date. Within VSOP members, subprojects are ongoing, such as developing the fully automatized VSOP Automatic Stellar Parameters (VASP), the VSOP Variability Classifier and the Binary Resolver. Then, papers are published by subgroups of VSOP \citep[The first VSOP paper is submitted, see][]{Dall-etal-2007}.}
\label{vsop_infrastructure}
\end{figure*}

\section{The VSOP Dataflow}
The VSOP dataflow comprises a collection of different steps that are developed in order to make it as automatic as possible. The overall VSOP infrastructure is outlined in Fig.~\ref{vsop_infrastructure}. Specifically: a) A list of targets is build based on the comparison between the GCVS and Simbad. b) When observing time is granted, we automatically generate one wiki-page per target. c) Observations for each target are defined and submitted to the observatory database. d) Observations are carried out through the observing semester, and data is automatically reduced by the instrument pipelines. e) Raw and reduced data are automatically transferred to the VSOP machine at ESO Vitacura and the wiki star page is updated. f) Plots of the cross-correlation function (CCF) are generated and included in the star pages. g) The stellar parameters are determined through the VSOP Automatic Spectral Parameters (VASP) software (not yet fully implemented). h) VSOP members receive an email alert that new VSOP observations have been obtained. i) Manual analysis and verification is undertaken, and the wiki-pages are updated.

The dataflow has proved very smooth and efficient throughout the first observing seasons. One still needs to develop an automatic variability classifier. This subproject of VSOP is currently ongoing, with collaboration with the Gaia and CoRoT teams \citep[see][]{Dall-etal-2007}.

\section{Extending the wiki philosophy: Wikimbad}

Based on the successful work model developed within V.S.O.P., and in particular the aspect of data sharing through a wiki website, one of us (C.F.) proposed to extend the philosophy of using a wiki for astronomical data to a larger scope. The project is called "Wikimbad" to illustrate both the fact that it is a wiki website, and the fact that it is strongly based on input provided by Simbad. It can be found at the following address: {\tt http://wikimbad.org}.

Wikimbad combines a few ideas in the same place, although none of these individual ideas alone is sufficient to justify the need for another astronomical service.
\begin{itemize}
\item "Being a Simbad for reduced data", in the sense of getting output based on a name request. Of course this exists already in large data archives, such as HST's. But for ESO's, data is not always reduced, and this possibility exists only for large institutional projects, but not for the very (numerous?) peculiar, old, personnal and/or small observational dataset.
\item Providing data reduced specifically by someone who then published these data, to the contrary of purely pipeline-reduced data (although some of the published data in the litterature is coming from pipeline outputs, but the real fraction of such data is yet unknown).
\item Providing reduced data specifically associated with a given publication, to open the possibility of a direct testing of the results published by someone else.
\item Allow to provide a very heterogeneous data package, not only the file formats (fits, ascii, data cubes), but also in the type of data (lightcurve, imaging, spectroscopy).
\item Be extremely simple and rapid to share the data, since no preparation whatsoever is required, nor any data characterization or standardization. The asociated publication plays the role of the essential validation. It thus avoids an large amounts of problems associated with the set up of the Virtual Observatory.
\item Provide the possibility for the datasets to be "discussed" (a feature of the MediaWiki software is the so-called "Talk" page asociated with every normal article page).
\end{itemize}

Wikimbad is up and running, and welcome any professional (or amator for that matter) astronomer to upload and share their own data, even if it has been published years ago. To keep the project flexible and managable, people are requested to register to the website before being allowed to upload and modify pages. Wikimbad cannot accept anonymous uploads.

Wikimbad hardly compare with the scope of the Virtual Observatory, since the latter is not only to provide a simple mean to find data, but also to compare them and start to analyze them. Moreover, Wikimbad is based purely on free\footnote{"Free" means here "free as in free speech" {\it and} "free as in free beer".} participation. However, Wikimbad has the straight advantage of existing right away and being extremely simple to use. The required association of the dataset with a publication ensure the dataset is actually described. This points is particularly sensitive, since it is a major concern of the VO. Moreover, the necessary credits to be given on a dataset if it is used in another publication is straightforward in Wikimbad, while it is also a difficulty in the VO.

\section{Conclusion: wiki and astronomy}

We have presented the Variable Star One-shot Project, and described its centralized website, based on the wiki technology. VSOP is a strong example (although others certainly exist) that a wiki is very well suited for professional collaboration between astronomers spread in different timezones. Wikimbad can be seen as a rather natural child project of the VSOP, although its scope is different, and is not restricted to a tight collaboration, but rather on personal involvement of astronomers by sharing their reduced and published data.

\section*{Acknowledgments}
The VSOP team reiterate its acknowledments to the IT departement of ESO Vitacura (in Chile) for material and software support. C.F. thanks LAOG director, J.-L. Monin, for accepting to host Wikimbad in the LAOG, and IT colleagues in LAOG, L. Torlay, F. Roussel and S. Lafrasse, for making this website a reality.


\begin{thebibliography}{3}
\expandafter\ifx\csname natexlab\endcsname\relax\def\natexlab#1{#1}\fi

\bibitem[{{Dall} {et~al.}(2007){Dall}, {Foellmi}, {Pritchard}, {Allende
  Prieto}, {Bruntt}, {Amado}, {Arentoft}, {Baes}, {Depagne}, {Fernandez},
  {Ivanov}, {Koesterke}, {Monaco}, {O'Brien}, M., {Saviane}, {Scharwaechter},
  {Schmidtobreick}, {Schuetz}, {Seifahrt}, {Selman}, {Stefanon}, \&
  {Sterzik}}]{Dall-etal-2007}
{Dall}, T.~H., {Foellmi}, C., {Pritchard}, J., {et~al.} 2007, \aap, submitted

\bibitem[{{Kazarovets} {et~al.}(2003){Kazarovets}, {Kireeva}, {Samus}, \&
  {Durlevich}}]{Kazarovets-etal-2003}
{Kazarovets}, E.~V., {Kireeva}, N.~N., {Samus}, N.~N., \& {Durlevich}, O.~V.
  2003, Informational Bulletin on Variable Stars, 5422, 1

\bibitem[{{Kholopov} {et~al.}(1998){Kholopov}, {Samus}, {Frolov}, {Goranskij},
  {Gorynya}, {Karitskaya}, {Kazarovets}, {Kireeva}, {Kukarkina}, {Kurochkin},
  {Medvedeva}, {Pastukhova}, {Perova}, {Rastorguev}, \&
  {Shugarov}}]{Kholopov-etal-1998}
{Kholopov}, P.~N., {Samus}, N.~N., {Frolov}, M.~S., {et~al.} 1998, in Combined
  General Catalogue of Variable Stars, 4.1 Ed (II/214A). (1998), 0--+

\end{thebibliography}
\end{document}